\documentclass[]{aastex63}

\usepackage{amsmath}

\usepackage{epigraph}
\received{}
\revised{}
\accepted{}
\submitjournal{PSJ}

\shorttitle{THAI Part 3}
\shortauthors{Fauchez et al.}
\graphicspath{{./}{figures_part3/}}
\begin{document}

\title{The TRAPPIST-1 Habitable Atmosphere Intercomparison (THAI). \\
Part III: Simulated Observables -- The return of the spectrum}

\author[0000-0002-5967-9631]{Thomas J. Fauchez}
\affiliation{NASA Goddard Space Flight Center, 8800 Greenbelt Road, Greenbelt, MD 20771, USA}
\affiliation{Goddard Earth Sciences Technology and Research (GESTAR), Universities Space Research Association (USRA), Columbia, MD 7178, USA}
\affiliation{American University, College of Arts and Sciences, Washington DC, USA}
\affiliation{NASA GSFC Sellers Exoplanet Environments Collaboration}

\author[0000-0002-2662-5776]{Geronimo L. Villanueva}
\affiliation{NASA Goddard Space Flight Center, 8800 Greenbelt Road, Greenbelt, MD 20771, USA}

\author[0000-0001-8832-5288]{Denis E. Sergeev}
\affiliation{Department of Mathematics,
College of Engineering, Mathematics, and Physical Sciences, University of Exeter
Exeter, EX4 4QF, UK}

\author[0000-0003-2260-9856]{Martin Turbet}
\affiliation{Observatoire  Astronomique  de  l’Universit\'e  de  Gen\`eve,  Universit\'e  de  Gen\`eve,  Chemin  des Maillettes 51, 1290 Versoix, Switzerland.}
\affiliation{Laboratoire de M\'et\'eorologie Dynamique/IPSL, CNRS, Sorbonne Universit\'e, \'Ecole Normale Sup\'erieure, PSL Research University, \'Ecole Polytechnique, 75005 Paris, France}

\author[0000-0002-1485-4475]{Ian A. Boutle}
\affiliation{Met Office, FitzRoy Road, Exeter, EX1 3PB, UK}
\affiliation{Department of Astrophysics, College of Engineering, Mathematics, and Physical Sciences, University of Exeter, Exeter, EX4 4QL, UK}

\author[0000-0001-5328-819X]{Kostas Tsigaridis}
\affiliation{Center for Climate Systems Research, Columbia University, New York, NY, USA}
\affiliation{NASA Goddard Institute for Space Studies, 2880 Broadway, New York, NY 10025, USA}

\author[0000-0003-3728-0475]{Michael J. Way}
\affiliation{NASA Goddard Institute for Space Studies, 2880 Broadway, New York, NY 10025, USA}
\affiliation{NASA GSFC Sellers Exoplanet Environments Collaboration}
\affiliation{Theoretical Astrophysics, Department of Physics and Astronomy, Uppsala University, Uppsala, Sweden}

\author[0000-0002-7188-1648]{Eric T. Wolf}
\affiliation{Laboratory for Atmospheric and Space Physics, University of Colorado Boulder, Boulder, CO, USA}
\affiliation{NASA NExSS Virtual Planetary Laboratory, Seattle, WA, 98195, USA}
\affiliation{NASA GSFC Sellers Exoplanet Environments Collaboration}

\author[0000-0003-0354-9325]{Shawn D. Domagal-Goldman}
\affiliation{NASA Goddard Space Flight Center, 8800 Greenbelt Road, Greenbelt, MD 20771, USA}
\affiliation{NASA GSFC Sellers Exoplanet Environments Collaboration}
\affiliation{NASA NExSS Virtual Planetary Laboratory, Seattle, WA, 98195, USA}

\author[0000-0002-3262-4366]{Fran\c cois Forget}
\affiliation{Laboratoire de M\'et\'eorologie Dynamique/IPSL, CNRS, Sorbonne Universit\'e, \'Ecole Normale Sup\'erieure, PSL Research University, \'Ecole Polytechnique, 75005 Paris, France}

\author[0000-0003-4346-2611]{Jacob Haqq-Misra}
\affiliation{NASA NExSS Virtual Planetary Laboratory, Seattle, WA, 98195, USA}
\affiliation{Blue Marble Space Institute of Science, Seattle, WA, USA}

\author[0000-0002-5893-2471]{Ravi K. Kopparapu}
\affiliation{NASA Goddard Space Flight Center, 8800 Greenbelt Road, Greenbelt, MD 20771, USA}
\affiliation{NASA GSFC Sellers Exoplanet Environments Collaboration}
\affiliation{NASA NExSS Virtual Planetary Laboratory, Seattle, WA, 98195, USA}

\author[0000-0003-4402-6811]{James Manners}
\affiliation{Met Office, FitzRoy Road, Exeter, EX1 3PB, UK}

\author[0000-0001-6707-4563]{Nathan J. Mayne}
\affiliation{Department of Astrophysics, College of Engineering, Mathematics, and Physical Sciences, University of Exeter, Exeter, EX4 4QL, UK}

\begin{abstract}
The TRAPPIST-1 Habitable Atmosphere Intercomparison (THAI) is a community project that aims to quantify how differences in general circulation models (GCMs) could impact the climate prediction for TRAPPIST-1e and, subsequently its atmospheric characterization in transit. Four GCMs have participated in THAI\deleted{ so far}: ExoCAM, LMD-Generic, ROCKE-3D and the UM.
This paper, focused on the simulated observations, is the third part of a trilogy, following the analysis of two land planet scenarios (Part I) and two aquaplanet scenarios (Part II). Here, we show a robust agreement between the simulated spectra and the number of transits estimated to detect the land planet atmospheres. For the \added{cloudy} aquaplanet ones,\replaced{ using atmospheric data from any of the four GCMs would require at least 17 transits}{a 5--$\sigma$ detection of CO$_2$ could be achieved in about 10 transits if the atmosphere contains at least 1 bar of CO$_2$}. \deleted{This prediction corresponds to  UM simulated data which produces the lowest and thinnest clouds.}   \replaced{Between 35-40$\%$}{That number can vary by 41--56$\%$ depending on the GCM used to predict the terminator profiles, principally due to differences in the cloud deck altitude, with ExoCAM and LMD-G producing higher clouds than ROCKE-3D and UM.} \deleted{more transits are predicted by ExoCAM or LMD-G due to higher thick terminator clouds.} \added{Therefore,} for the first time, this work  provides ``GCM uncertainty error bars'' of \replaced{$\sim$35-40$\%$}{$\sim 50\%$} that need to be considered in future analyses of transmission spectra. We also analyzed the inter-transit \added{spectral} variability \deleted{induced by weather patterns and changes of terminator cloudiness between transits}. Its magnitude differs significantly between the GCMs but its impact on the transmission spectra is within the measurement uncertainties.  THAI has demonstrated the importance of model intercomparison for exoplanets and also paved the way for a larger project to develop an intercomparison meta-framework, namely the Climates Using Interactive Suites of Intercomparisons Nested for Exoplanet Studies (CUISINES).
\end{abstract}

%% Keywords should appear after the \end{abstract} command. 
%% See the online documentation for the full list of available subject
%% keywords and the rules for their use.

%% From the front matter, we move on to the body of the paper.
%% Sections are demarcated by \section and \subsection, respectively.
%% Observe the use of the LaTeX \label
%% command after the \subsection to give a symbolic KEY to the
%% subsection for cross-referencing in a \ref command.
%% You can use LaTeX's \ref and \label commands to keep track of
%% cross-references to sections, equations, tables, and figures.
%% That way, if you change the order of any elements, LaTeX will
%% automatically renumber them.
%%
%% We recommend that authors also use the natbib \citep
%% and \citet commands to identify citations.  The citations are
%% tied to the reference list via symbolic KEYs. The KEY corresponds
%% to the KEY in the \bibitem in the reference list below. 

\section{Introduction} \label{sec:intro}
\epigraph{Here at last... comes the end of our fellowship. I will not say do not weep, for not all tears are an evil.}{---J.R.R. Tolkien, \textit{The Return of the King}}

At the dawn of terrestrial exoplanet atmospheric characterization with the James Webb Space Telescope (JWST), predicting the detectability of the atmospheres of such planets is crucial in order to prepare for observations and maximize the scientific return. JWST Guaranteed Time Observations (GTO) and Cycle 1 proposals have already been selected. \replaced{but}{While CO$_2$ can be potentially detectable from Cycle 1,} it is unlikely that enough transits would be accumulated for any single target \deleted{within cycle 1} to characterize \added{in-depth} \deleted{or even detect} the atmosphere of a terrestrial exoplanet \citep{Fauchez:2019,Lustig_Yaeger2019,Pidhorodetska_2020} in the Habitable Zone  \citep[HZ; see e.g.][]{Kopparapu2013} of M--dwarf stars.
The presence of CO$_2$ has been shown to be the best proxy \citep{Fauchez:2019,Lustig_Yaeger2019,Turbet:2020trappist1review} for the detection of a potentially habitable atmosphere due to its strong absorption band in the mid-infrared (MIR) at 4.3~$\mu m$ and in the far-infrared at $\sim$ 15~$\mu m$. However a 5--$\sigma$ detection under cloudy conditions would likely require more than a dozen transits, even for the most favorable HZ terrestrial planet, TRAPPIST-1e \citep{Fauchez:2019}.\\

TRAPPIST-1e belongs to the seven small transiting planet system TRAPPIST-1 \citep{Gillon2016,Gillon2017,Luger2017trappist} at 12.0~pc away. The star, TRAPPIST-1, is a M8V just slightly larger than Jupiter which makes it very suitable for transmission spectroscopy of the atmosphere of small rocky planets. Indeed, the ratio of the surface area of the star's disk blocked out by the planet's disk (including its atmosphere), i.e. the transit depth, is inversely proportional to the square of the star radius. Also, around such \added{a} cold and dim star\added{,} HZ planets have a very short orbital period leading to very frequent transits and therefore more accessible data on their atmospheres.
The Hubble Space Telescope (HST) has been used  to infer the presence of an atmosphere on TRAPPIST-1e and its sibling planets \citep{deWit2016,deWit2018}. However, the precision of HST data were only able to \added{either be consistent with the absence of an atmosphere or to} rule out clear--sky H$_2$ dominated atmospheres while \cite{Moran2018} showed that HST observations could actually be fit by cloudy/hazy H$_2$ atmospheres. Yet, \added{the comparison of} TRAPPIST-1e bulk density measurements \citep{Grimm18,Agol2021} \replaced{comparison with}{to} H$_2$-rich planets mass-radius relationships \citep{Turbet:2020trappist1review} along with atmospheric escape modelling and gas accretion modelling \citep{hori:2020} provide accumulating evidence against the presence of H$_2$ dominated cloudy atmospheres around TRAPPIST-1 planets, including TRAPPIST-1e (see \citealt{Turbet:2020trappist1review} and references therein). Furthermore, \citet{Krishnamurthy2021} reported strong upper limit constraints on the absence of helium in the atmosphere of TRAPPIST-1e. More in-depth knowledge about the absence or presence of a high mean molecular weight atmosphere on TRAPPIST-1e would most likely require JWST transit observations. Indeed, even some of the largest planned optical telescope (Extremely Large Telescopes, ELTs) are unable --- even at the diffraction limit --- to separate the light from TRAPPIST-1 and its planets, because of their very small angular separation. The same is true for future space observatories such as the Roman \citep{Douglas_2020}, Large UV/Optical/Infrared Surveyor (LUVOIR, \cite{team2019luvoir}), the Habitable Exoplanet Observatory HabEx (\cite{HabEx2018}) for which the inner working angle of their coronagraph  would block the light not only from the star but from the entire system. Also, in the HZ, the planet is relatively too cold to significantly emit thermal infrared radiation which makes it very challenging to characterize its emission spectrum \citep{Fauchez:2019,Lustig_Yaeger2019,Kane2021}; more close-in planets, however, will be more sensitive to this technique \citep{Morley2017,Lustig_Yaeger2019,Koll2019,Turbet:2020trappist1review}.
Orbital broadband phase-dependent variations in the combined planetary thermal emission and reflected stellar energy can also provide clues about the atmospheric structure and surface properties of the planet \citep[e.g.][]{Selsis2011,vonparis2016,Koll_Abbot:2016,Turbet:2016,Haqq_Misra2018,Wolf2019}, but the level of the phase variation of part-per-billion (ppb) are far beyond our current instrumental capabilities \citep{Wolf2019}. Atmospheric characterization of TRAPPIST-1e may be also possible from the ground \citep{Wunderlich2020} with the planned European Extremely Large Telescope (E-ELT), but this demonstration has been done assuming a clear-sky TRAPPIST-1e atmosphere, which may have led to an over-estimation of the E-ELT's capabilities.

Therefore, JWST transit observations are the most viable atmospheric characterization technique for the TRAPPIST-1 planets in the coming decade. Several studies have used either GCMs \citep{Fauchez:2019,Pidhorodetska_2020,May2021water} or 1-D radiative convective climate models coupled to photochemistry \citep{Lincowski2018,Lustig_Yaeger2019,Wunderlich2019,Wunderlich2020,Lin2021} or analytical models \citep{Morley2017} to predict the detectability of standard atmospheres such as the modern Earth, the Archean Earth or a CO$_2$-dominated atmosphere. While these predictions inherently vary from one model category to another due to for instance the day/night contrast or the presence of clouds and hazes in the simulated atmosphere, models in the same category may also disagree due to, for example, differences in the atmospheric profiles at the terminator.
Evaluating these differences and their impact on synthetic observations in order to optimize JWST observation strategies are the core objectives of the TRAPPIST-1 Habitable Atmosphere Intercomparison \citep[THAI;][]{Fauchez2020THAI,Fauchez2021_THAI_workshop}.  The comparison of these simulated climate systems are described in the companion papers \citep[][referred to as Part I]{Turbet21_THAI} for the dry planet benchmark scenarios (Ben~1 \& Ben~2) and in \citep[][referred to as Part II]{Sergeev21_THAI} for the aquaplanet (Hab~1 \& Hab~2) scenarios. The paper is structured as follows: In Section~\ref{sec:method} the methods and tools used in \added{t}his study are described. In Section~\ref{sec:transmi}, we present the simulated transmission spectra using each of the GCM outputs, using both time average and time dependent terminator profiles. Finally, conclusions are given in Section~\ref{sec:conclusions}.

\section{Method} \label{sec:method}
\subsection{The THAI GCM simulations}
In this paper, Part III of a trilogy of THAI papers, we use the same GCM simulations that have been extensively analyzed in Part I \citep{Turbet21_THAI} and Part II \citep{Sergeev21_THAI}, namely the Ben~1 \& Ben~2 and Hab~1 \& Hab~2 cases, respectively. Briefly, the Ben~1 \& Ben~2 cases are dry land planet simulations, while the Hab~1 \& Hab~2 cases assume that the surface is fully covered by a global (static) ocean and that there is water vapour and clouds in the atmosphere. Ben~1 \& Hab~1 atmospheric composition is broadly similar to that of modern Earth (1 bar of N$_2$, 400 ppmv of CO$_2$), while the Ben~2 \& Hab~2 experiments assume a CO$_2$-dominated atmosphere (1~bar). 10 orbits (61 Earth days) at a frequency of 6~h are output for Ben~1 \& Ben~2, while 100 orbits (610 Earth days) are output for Hab~1 \& Hab~2, in order to smooth out the internal variability with a period of about a dozen of orbits induced by clouds.

Each of these simulations have been performed by the four THAI GCMs: ExoCAM \citep{Wolf2022} the exoplanet branch of the Community Earth System Model (CESM, \url{http://www.cesm.ucar.edu/models/cesm1.2/}) version 1.2.1., the LMD Generic model \citep[LMD-G, see e.g.][]{Wordsworth:2011,Turbet:2018aa}, the Resolving Orbital and Climate Keys of Earth and Extraterrestrial Environments with Dynamics \citep[ROCKE-3D][]{Way2017}) and the Met Office Unified Model \citep[the UM, see e.g.][]{Mayne_2014b,Boutle2017}. More details on these four GCMs are also provided in the THAI protocol \citep{Turbet21_THAI,Sergeev21_THAI} and the THAI workshop report \citep{Fauchez2021_THAI_workshop}.

\subsection{Simulated Spectra}
We use the Planetary Spectrum Generator \citep[PSG,][]{Villanueva2018,Villanueva2022} to simulate transmission spectra of TRAPPIST-1e for each of the THAI scenarios, using data from each of the four GCMs. PSG is an online radiative transfer tool that can be used to simulate planetary spectra observations from any ground or space-based observatory for various objects of the solar system and beyond, and it also includes a noise calculator.

To simulate and compare time-averaged transmission spectra across the models, we first average the atmospheric properties over the 10 orbits of Ben~1 \& Ben~2 and over the 100 orbits of Hab~1 \& Hab~2. \replaced{PSG then computes the transmission spectrum from each of the latitude grid boxes around the terminator.}{
For our transit calculations, atmospheric profiles were created for each GCM latitude $\times$ longitude cell at the terminator of the planet using abundance, pressures, temperatures as reported by the GCM for that specific terminator cell (vertical parameters of that cell). Then transit spectra were computed using those profiles across all terminator cells, and the total transit spectrum was computed by the average of all transits across the terminators (weighted by the latitudinal extension of the cell). }

\replaced{Molecular opacities \replaced{, including}{and} homonuclear and heteronuclear collision induced absorptions (CIAs) are taken from the HITRAN \citep{Gordon:2022} \added{and \citep{Karman:2019}, respectively}\deleted{, updated with the 2020 version}.}{
Specifically, the radiative transfer is computed employing a layer-by-layer pseudo-spherical refractive raytracing algorithm. Rayleigh cross-sections are computed as a summation of the individual molecular cross-sections \citep{Villanueva2022,Sneep2005}, which are computed at each wavelength based on the polarizability of the encompassing molecules. PSG employs polarizability values as compiled on the Computational Chemistry Comparison and Benchmark DataBase at NIST (\url{https://cccbdb.nist.gov/pollistx.asp}). Collision-induced absorptions (CIA) generated by inelastic collisions of molecules in a gas are included from considering the latest HITRAN CIA compilations \citep{Gordon:2022} and and the HITRAN CIA database \citep{Karman:2019}, as well as several other sources as reported in \citep{Villanueva2022}, including the MT\_CKD water continuum  \citep{Mlawer2012}, here in version v3.5 \citep{Payne2020,Kofman2021}). In the presented spectral range and assumed background abundance, only these CIAs contain notable signatures: CO$_2$--CO$_2$, H$_2$O--H$_2$O, H$_2$O--N$_2$, N$_2$--N$_2$. Molecular absorptions were included via correlated-k tables based on the latest HITRAN 2020 linelist \citep{Gordon:2022}, which were complemented at short wavelengths ($<1$ $\mu m$) with UV cross-sections, primarily from the MPI Mainz UV/VIS Spectral Atlas \citep{KellerRudek:2013}.} Aerosols properties are modelled following Mie theory, with water cloud scattering properties as described in \cite{MassieHervig:2013} while the ice cloud optical property parameterization uses \cite{Warren1984}, as also described in \cite{MassieHervig:2013}. The partial abundance of the aerosols at the grid-box is explicitly defined by the GCM kg/kg profile at that location, meaning that a profile with zero abundance would correspond to a fully clear scenario. An average of all these spectra is then computed to obtain a limb-averaged spectra as it would be observed by an instrument. \added{Detailed information regarding the computation of correlated-k tables, Rayleigh scattering, the treatment of CIAs, MT\_CKD, the ray tracing algorithm, and the radiative-transfer method can be found in \citep{Villanueva2022}.}

\deleted{As stated in the Sec.~\ref{sec:intro}, TRAPPIST-1e is too cold to be characterized via emission spectroscopy \citep{Morley2017,Fauchez:2019,Lustig_Yaeger2019} and it is at far too small angular separation from its star to be accessible to direct imaging techniques, so these techniques are not considered here.}

\subsection{Instruments, noise and number of transits}
\subsubsection{Instruments}
We have simulated JWST observations of TRAPPIST-1e transiting its host star. JWST is a 6.5~m tip-to-tip segmented telescope, equivalent to a full circular aperture of 5.6~m of diameter. Previous studies have showed that the NIRSpec Prism (covering the $0.6-5.3~
\mu m$ region at resolving power R=\replaced{300}{100}) is the JWST instrument most adapted to characterize the atmosphere of temperate terrestrial planets \citep{Fauchez:2019, Lustig_Yaeger2019,Pidhorodetska_2020,Wunderlich2020}. Indeed, this spectral region contains various molecular signatures of interest such as O$_2$, \added{O$_3$,} H$_2$O, NO$_2$, N$_2$O, CH$_4$, CO and CO$_2$. The latter may likely be the only one with a strong enough absorption features \deleted{(at 4.3 $\mu m$)} to be detectable with JWST in a reasonable number of transits \added{(i.e. achievable in 5 JWST cycles assuming a constant 4 transit observation per cycle as in Cycle 1)}, when clouds and hazes are present in the atmosphere. It has therefore been suggested as the best proxy to detect the atmosphere of habitable planets \citep{Fauchez:2019,Turbet:2020trappist1review}. Note that JWST \deleted{General Observer (GO) and} Guaranteed Time Observations (GTO) proposals have \replaced{also}{already} been awarded for the \replaced{NIRISS and MIRI}{NIRSpec} instrument\deleted{s} that will attempt molecular detection \added{with four transit observations (program 1331)}.
In this work we compute spectra from 0.6 to 20 $\mu m$, across the range of both JWST NIRSpec Prism and MIRI medium-resolution spectrometer (MRS) and present the figures at R=100 offering the best visibility for multiple spectrum plots. \added{Signal-to-noise ratios (S/N) and number of transit estimations are only presented for NIRSpec Prism}.

\subsubsection{Noise}\label{subsub:noise}
\subsubsection{Estimation of the detectability of the atmosphere --- identical transits}
\deleted{The number of transits required to detect an atmosphere is calculated using the CO$_2$ line at 4.3 $\mu m$ as proxy. In the presence of N$_2$ in the atmosphere, this band overlaps with the N$_2$-N$_2$ collision-induced absorption (CIA).}

\added{First,} we computed noise estimates with PSG, and validated these by employing the official JWST Exposure Time Calculator, obtaining very good agreement. For NIRSpec Prism, the effective spectral resolution is 0.022~$\mu m$\deleted{corresponding to a resolving power of 195 near 4.3 $\mu m$}. We have selected the clear filter with the sub--array SUB512S and the rapid readout pattern with two groups per integration and 0.225~s per frame. This leads to a partial saturation near the peak of the Stellar Energy Distribution (SED) following \citep{Batalha:2018,Lustig_Yaeger2019}\deleted{ but not in the region of interest (4--5$\mu m$)}. For MIRI LRS, the effective resolution is 
0.0654 ~$\mu m$, and we selected the P750L disperser with the sub--array SLITLESSPRISM, and a FASTR1 readout pattern with 20 groups per integration with a frame time of 0.15~s. \\
\added{We assumed a transit time of 3345~s (0.93~h) \citep{Agol2021}. To take into account the noise of the out-of-transit baseline, we used JWST NIRSpec GTO proposal 1331 \citep{Lewis_2018} for which each transit event will be observed for $\sim$4 hours, therefore leading to an out-of-transit observation of $\sim$3$\times$ transit duration. As the noise adds in quadrature, the single transit noise $N$, including a 3$\times$ out-of-transit baseline is computed as follow:} 
\begin{equation}
    \begin{aligned}
    N & = \sqrt{N_{out}^2 + N_{in}^2}\\
      & = \sqrt{(1/time_{out})^2 + (1/time_{in})^2}\\
      & = \sqrt{(1/3)^2 + (1/1)^2}\\
      & = \sqrt{4/3}
    \end{aligned}
\end{equation}

\replaced{To estimate at which S/N we can detect an atmosphere using the 4.3 $\mu m$ feature as a proxy we subtract the spectrum with CO$_2$ (and N$_2$ when present)  by the one without CO$_2$ nor N$_2$ and we divide by the noise. The S/N of the band is then computed following \replaced{\cite{Kaltenegger2009}}{\cite{Lustig_Yaeger2019}} as:

\begin{equation*}
    S/N = \sqrt{\sum_{i=0}^{n}S/Ni^2},
\end{equation*}
where S/Ni are the individual S/N in each spectral interval within the 4-4.6 $\mu m$ region encompassing the CO$_2$ band.

From the S/N, the number of transits $N_t$ required to achieve a 5--$\sigma$ detection is done as following \citet{Fauchez:2019}:
\begin{equation*}
    N_t = \left(\frac{5}{S/N}\right)^2.
\end{equation*}
}{
To estimate the 1 transit S/N of CO$_2$ across the NIRSpec Prism range (0.6-5.3~$\mu m$) and the number of transits required to achieve a 5--$\sigma$ detection of CO$_2$ we proceed following the list below:
\begin{enumerate}
    \item We compute the spectrum without CO$_2$ (but keeping the other gases in)
    \item We compute the spectrum with CO$_2$
    \item We compute the difference between step 1. and 2 across the whole instrument range
    \item We compute the S/N by dividing step 3. by the noise for 1 transit (3345~s) in each spectral interval.
    \item We apply the $\sqrt{4/3}=1.155$ factor to the noise to take into account the out-of-transit noise.
    \item  The S/N of CO$_2$ across the whole instrument range  then computed following \replaced{\cite{Kaltenegger2009}}{\cite{Lustig_Yaeger2019}} as:
    \begin{equation}
    S/N = \sqrt{\sum_{i=0}^{n}S/Ni^2},
    \end{equation}
    where S/Ni are the individual S/N in each spectral interval.
    \item From the S/N, the number of transits $N_t$ required to achieve a 5--$\sigma$ detection is done as following \citet{Fauchez:2019,Fauchez2020O2}:
\begin{equation}
    N_t = \left(\frac{5}{S/N}\right)^2.
\end{equation}
\end{enumerate}
}
\section{Transmission Spectra}\label{sec:transmi}
\subsection{Ben~1 \& Ben~2 cases (dry planets)}
\begin{figure}[h]
\centering
\resizebox{15cm}{!}{\includegraphics{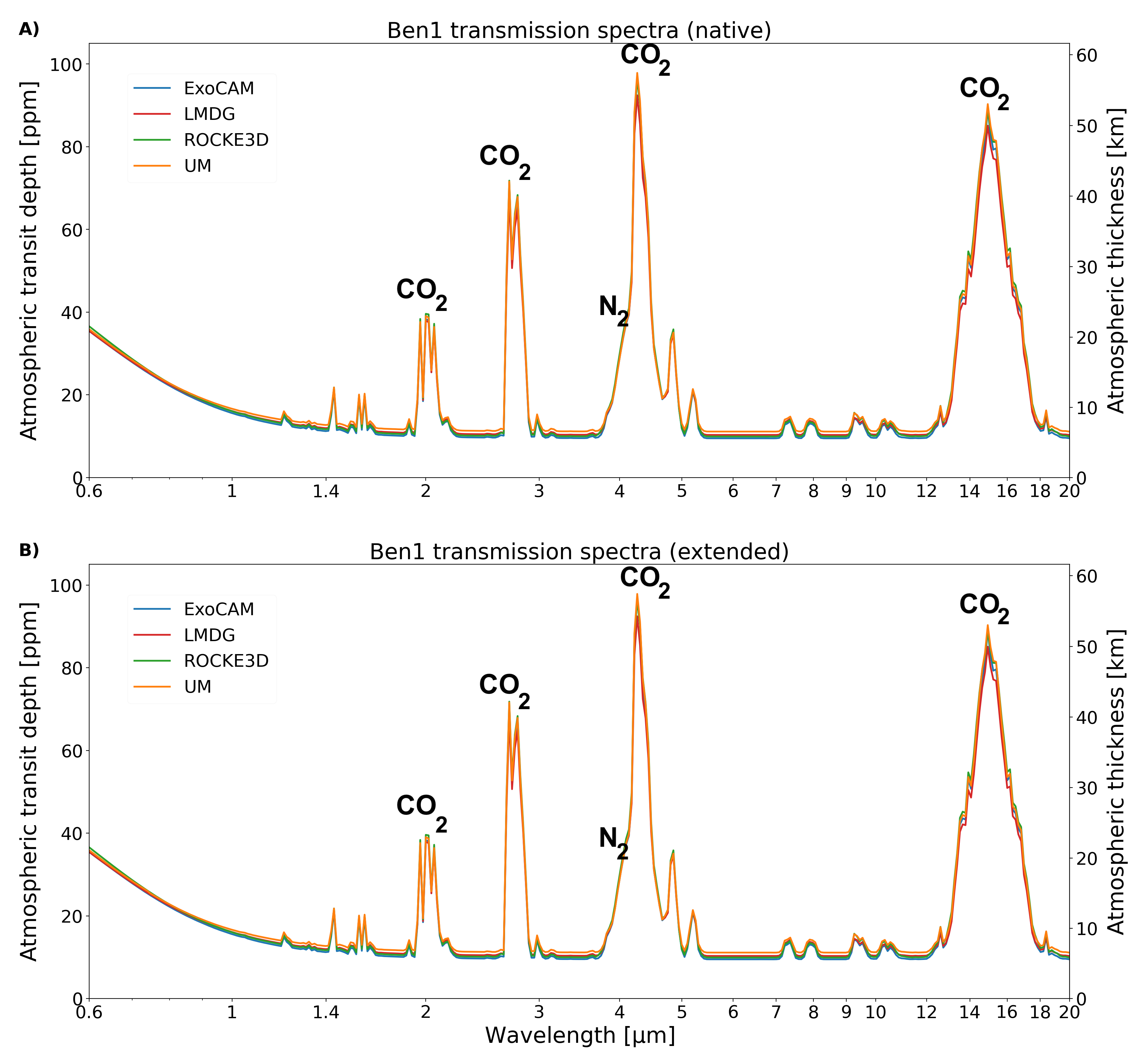}}
\caption{Ben~1 transmission spectra simulated with PSG using the terminator atmospheric profiles from the ExoCAM, LMD-G, ROCKE-3D and the UM simulations. In Panel A) the atmospheric profiles are limited up to the lowest pressure at the top of the GCMs, namely 10$^{-5}$, 4$\times$10$^{-5}$, 14$\times$10$^{-5}$ and 4$\times$10$^{-5}$~bar for ExoCAM, LMD-G, ROCKE-3D and UM, respectively while in Panel B) the atmospheric profiles have been extended up to 10$^{-7}$~bar assuming an isothermal atmosphere and fixed mixing ratios for the dry gases. The differences between the transmission spectra are extremely small.}
\label{Fig:Ben1}
\end{figure}

\begin{figure}[h]
\centering
\resizebox{15cm}{!}{\includegraphics{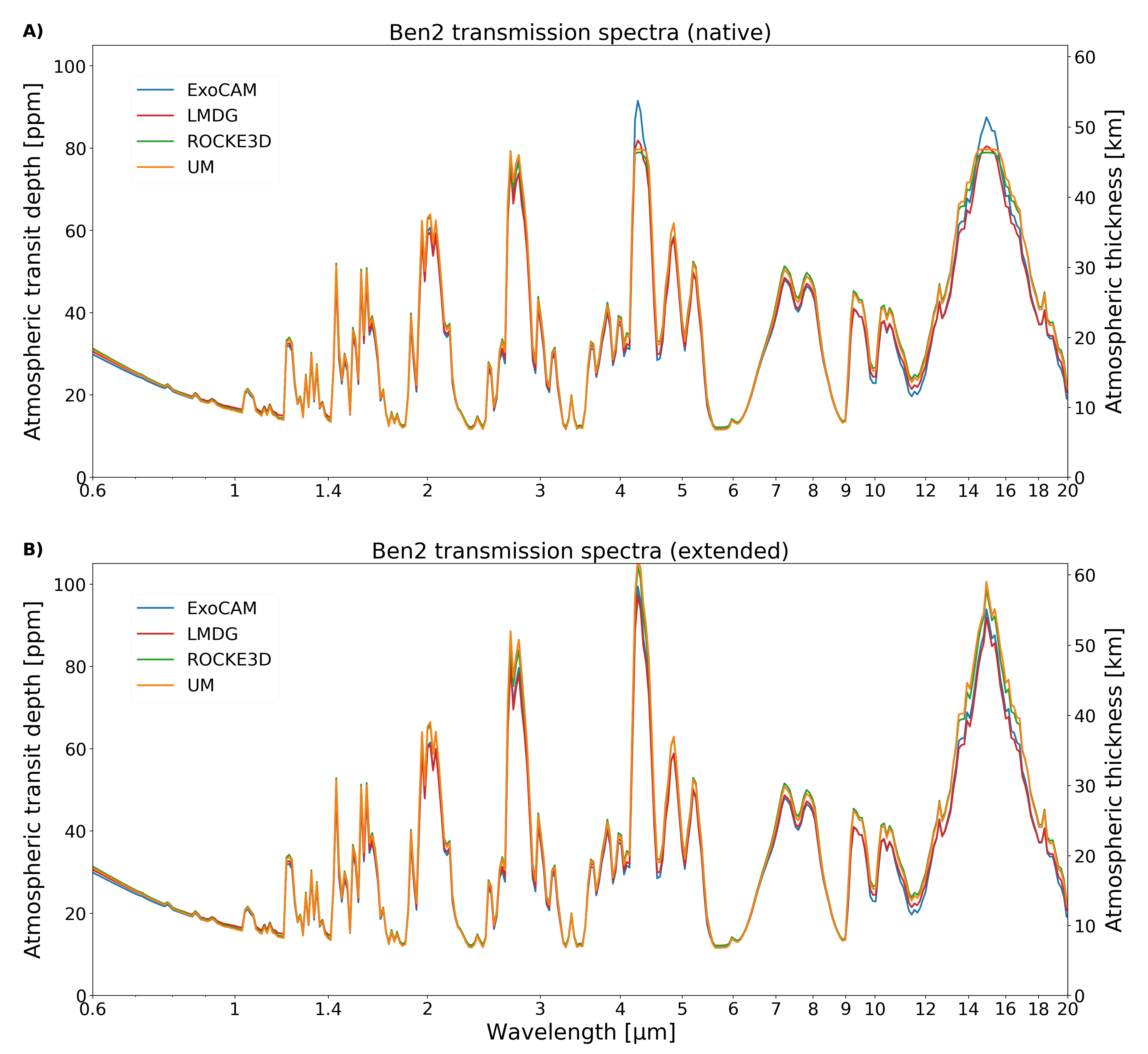}}
\caption{Ben~2 transmission spectra simulated with PSG using the terminator atmospheric profiles from the ExoCAM, LMD-G, ROCKE-3D and UM simulations. In Panel A) the atmospheric profiles are limited up to the lowest pressure at the top of the GCMs, namely 10$^{-5}$, 4$\times$10$^{-5}$, 14$\times$10$^{-5}$ and 13$\times$10$^{-5}$~bar for ExoCAM, LMD-G, ROCKE-3D and the UM, respectively  while in Panel B) the atmospheric profiles have been extended up to 10$^{-10}$ bar assuming an isothermal atmosphere and fixed mixing ratios for the dry gases. CO$_2$ features are truncated if the atmospheres are not vertically extended. Visible differences between the spectra appear in the CO$_2$ bands in both the native (panel A) and extended (panel B) scenarios. \added{CO$_2$ labels are not shown as the spectra are exclusively CO$_2$.}}
\label{Fig:Ben2}
\end{figure}

In Fig.~\ref{Fig:Ben1} and \ref{Fig:Ben2} we can see the transmission spectra simulated with PSG using the atmospheric profiles of THAI cases Ben~1 \& Ben~2, respectively, provided by each GCM. 
In the panel A) of both figures, the lowest pressures (highest altitudes) used to compute the spectra correspond to the top of the modelled domains, which are 10$^{-5}$, 4$\times$10$^{-5}$, 14$\times$10$^{-5}$ and 4$\times$10$^{-5}$~bar for Ben~1 and 10$^{-5}$, 4$\times$10$^{-5}$, 14$\times$10$^{-5}$ and 13$\times$10$^{-5}$~bar for Ben~2 for ExoCAM, LMD-G, ROCKE-3D and the UM, respectively. Note that most GCMs have the domain lid at relatively high pressures for numerical stability reasons and as moving it to lower pressures require taking into account complex upper atmospheric processes such as non-local thermodynamic equilibrium,  molecular diffusion, etc (see \cite{Fauchez2021_THAI_workshop}, section 4.1). The lowest pressures usually correspond to a top-of-atmosphere (TOA) altitude of 50 to 70~km for Earth-like simulations. However, the pressure at this altitude is usually too high to fully capture the transmitted light through the planet's atmosphere and the strongest atmospheric features can be truncated. This is clearly seen in the Ben~2 case where the CO$_2$ strongest absorption lines are truncated (Fig.~\ref{Fig:Ben2}a). To bypass this limitation, we used PSG to extend the atmosphere to much lower pressures, assuming an isothermal profile and constant volume mixing ratios for the dry gases. This is similar to the so-called ``ghost layer'' used in \cite{Amundsen_2016}. We have estimated the TOA pressure that would fully resolve the spectral lines for Ben~1 \& Ben~2  as 10$^{-7}$ and 10$^{-10}$~bar, respectively. Lower pressures are required for Ben~2 because the opacity of a pure CO$_2$ atmosphere remains strong at lower pressures than that of a N$_2$-dominated atmosphere.

Using the data with extrapolated model top, we have estimated the number of transits that would be required to detect such atmospheres with a 5--$\sigma$ confidence level \added{as presented in Table \ref{tab:transit}. Two estimations are shown, the first one uses the method presented in section \ref{subsub:noise} and is referred to as 5$\sigma$ Transits. The second method only uses the CO$_2$ line at 4.3$~\mu m$ as done in  \citep{Fauchez:2019,Wunderlich2019,Pidhorodetska_2020} and is referred to as 5$\sigma$ Transit-4.3$\mu m$. Also shown are the signal-to-noise ratio for 1 transit  (S/N-1) and four transits (S/N-1), corresponding to JWST Cycle 1 TRAPPIST-1e transit observation \citep[GTO Proposal 1331 by PI Nikole Lewis,][]{LewisT1eJWST}).}

\replaced{For Ben~1 we found the minimum number of transits to be 17, 19, 17 and 17 for ExoCAM, LMD-G, ROCKE-3D and the UM, respectively; for Ben~2 the numbers are 17, 21, 17 and 18.}{First, we can see that for a Ben~1 atmosphere, an average of $2.6~\sigma$ could be achieved from Cycle 1 while an average of $4.3~\sigma$ could be achieved for a Ben~2 atmosphere with more CO$_2$. To reach the necessary 5--$\sigma$ threshold, an average of 17 and 6 transits would be needed for Ben~1 and Ben~2, respectively. When using only the CO$_2$ line at 4.3$\mu m$ these numbers go up to 24 and 25 transits, respectively. This demonstrates that this method strongly over-estimates the number of necessary transits, especially if the amount of CO$_2$, and therefore the number of strong lines, is high.} The inter-model differences \added{for 5--$\sigma$ Transit} are small in both cases \added{, the maximum difference between the GCMs being 24 $\%$ and 33 $\%$ for Ben~1 and Ben~2, respectively} demonstrating that the four GCMs provide similar atmospheric profiles at the terminator to provide consistent simulated spectra and expected number of transits to detect a dry 1~bar atmosphere with a relatively high mean molecular weight. Note that we did not consider the spectral impact of dust that can be lifted from the surface of a land planet and persist in the atmosphere. Dust would likely raise the continuum level, thereby decreasing the amplitude of each spectral line as shown in \citet{Boutle_2020}. The expected effect would be of the order of 10~ppm.

\begin{table}
\centering
\caption{Table summarizing the signal-to-noise ratio achieved from 1 transit (S/N-1), from the four JWST Cycle 1 transits (S/N-4), number of transits to reach a 5--$\sigma$ detection using all available CO$_2$ lines (5--$\sigma$ Transit), and using only the 4.3 $\mu m$ CO$_2$ line.} \label{tab:transit}
\begin{tabular}{c c c c c}
\hline
\hline
  & \multicolumn{4}{c}{\bf{Ben1}} \\
Model & S/N-1  & S/N-4 & 5--$\sigma$ Transit & 5--$\sigma$ Transit-4.3$\mu m$  \\
ExoCAM & 1.3  & 2.6  & 16 &  23 \\
LMD-G & 1.2 &  2.4 & 19 & 25  \\
ROCKE-3D &  1.3& 2.6 & 15 & 23  \\
UM & 1.3 & 2.6  & 16 & 23  \\
Average & 1.3 & 2.6 & 17 & 24  \\
Maximum difference (\%)& 8 & 8 & 24 & 8  \\
\hline

  & \multicolumn{4}{c}{\bf{Ben2}} \\
Model & S/N-1  & S/N-4 & 5--$\sigma$ Transit & 5--$\sigma$ Transit-4.3$\mu m$  \\
ExoCAM & 2.2  & 4.4  & 5 & 23  \\
LMD-G & 2.0  &  4.0 &7 & 28\\
ROCKE-3D & 2.2  & 4.4 & 5 & 23  \\
UM & 2.2 & 4.4 &  5  & 24 \\
Average & 2.2  & 4.3 &  6 & 25 \\
Maximum difference (\%)& 9 & 9 & 33 & 20 \\
\hline

  & \multicolumn{4}{c}{\bf{Hab1}} \\
Model & S/N-1  & S/N-4 & 5--$\sigma$ Transit & 5--$\sigma$ Transit-4.3$\mu m$  \\
ExoCAM & 0.9 & 1.8   & 35 & 39 \\
LMD-G & 0.9 & 1.8  & 29 & 37 \\
ROCKE-3D & 1.0  & 2.0 & 38 &  31 \\
UM & 1.0 &  2.0 & 23 &  24 \\
Average & 1.0 & 2.0 & 29  & 33  \\
Maximum difference (\%)& 10 & 10 & 41 & 45 \\
\hline

  & \multicolumn{4}{c}{\bf{Hab2}} \\
Model & S/N-1  & S/N-4 & 5--$\sigma$ Transit & 5--$\sigma$ Transit-4.3$\mu m$  \\
ExoCAM & 1.5  & 3.0 & 12 & 36  \\
LMD-G & 1.7 & 3.4  & 9 & 31 \\
ROCKE-3D & 1.7 & 3.4 & 8 & 29 \\
UM & 2.0 & 4.0  & 7 &  23\\
Average & 1.7 & 3.4 & 9 &  30 \\
Maximum difference (\%)& 29 & 29 & 56 & 43 \\

\hline
\hline
\end{tabular}
\end{table}

\subsection{Hab~1 \& Hab~2 cases (aquaplanets)}
Rocky exoplanets in the HZ and with surface liquid water will likely have water in a vapor and condensed form in the atmosphere. Clouds have been shown to severely impede atmospheric characterization via transmission spectroscopy \citep{Fauchez:2019,Komacek2020,Suissa2020}. Furthermore, clouds are notoriously difficult to represent correctly in GCMs, because the characteristic timescale and size of individual clouds is too small to be simulated explicitly and they involve\deleted{d} a tremendous amount of physical processes. GCMs thus rely on sub-grid scale parameterizations to represent the formation of clouds that can significantly differ between models \citep{Sergeev21_THAI}. These discrepancies can then lead to different predicted surface temperatures, as was noted in exoplanet GCM simulations by \citet{Yang2019}. Details on the differences in GCM predictions, especially for clouds, are given in the companion paper \citep{Sergeev21_THAI}. \\
Here, we use PSG to compute transmission spectra for both the Hab~1 \& Hab~2 cases. The atmospheric properties at the terminator have been time averaged over the 100 orbits in order to smooth out variability that can be introduced by weather patterns and change in clouds at the terminator, as is commonly done for such planets \citep{Fauchez:2019,Komacek2020,Pidhorodetska_2020,Suissa2020,Suissa2020b}. Details on the impact of atmospheric variability on transmission spectra are given in Sec.~\ref{subsec:variability}.

\begin{figure}[h]
\centering
\resizebox{15cm}{!}{\includegraphics{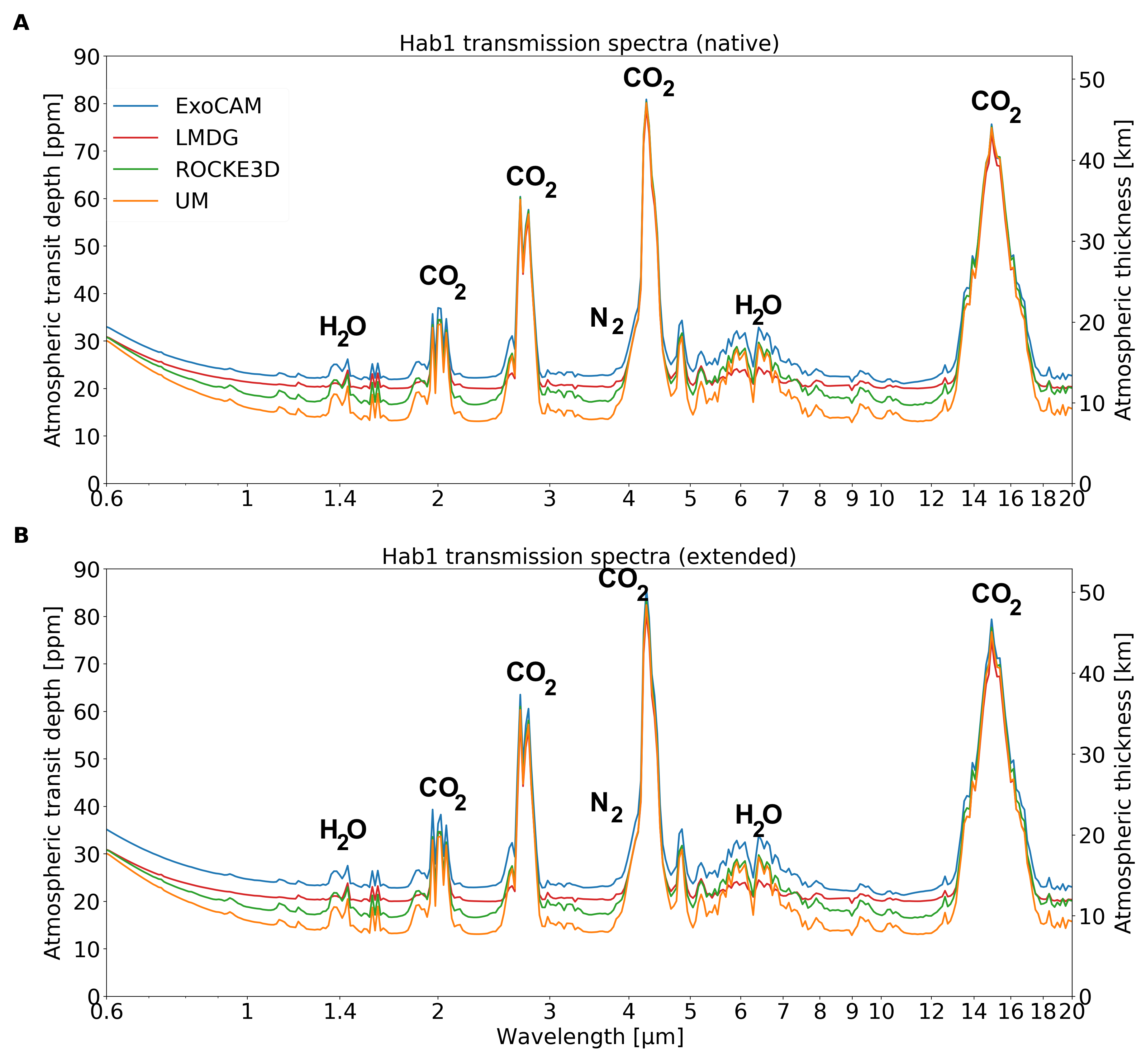}}
\caption{Hab~1 transmission spectra simulated with PSG using the terminator atmospheric profiles from the ExoCAM, LMD-G, ROCKE-3D and the UM simulations. In Panel A) the atmospheric profiles are limited up to the lowest pressure a\added{t} the top of the GCMs, namely 10$^{-5}$, 4$\times$10$^{-5}$, 14$\times$10$^{-5}$ and 4$\times$10$^{-5}$~bar for ExoCAM, LMD-G, ROCKE-3D and the UM, respectively  while in Panel B) the atmospheric profiles have been extended up to 10$^{-7}$ bar assuming an isothermal atmosphere and fixed mixing ratios for the dry gases.}
\label{Fig:Hab1}
\end{figure}

\begin{figure}[h]
\centering
\resizebox{15cm}{!}{\includegraphics{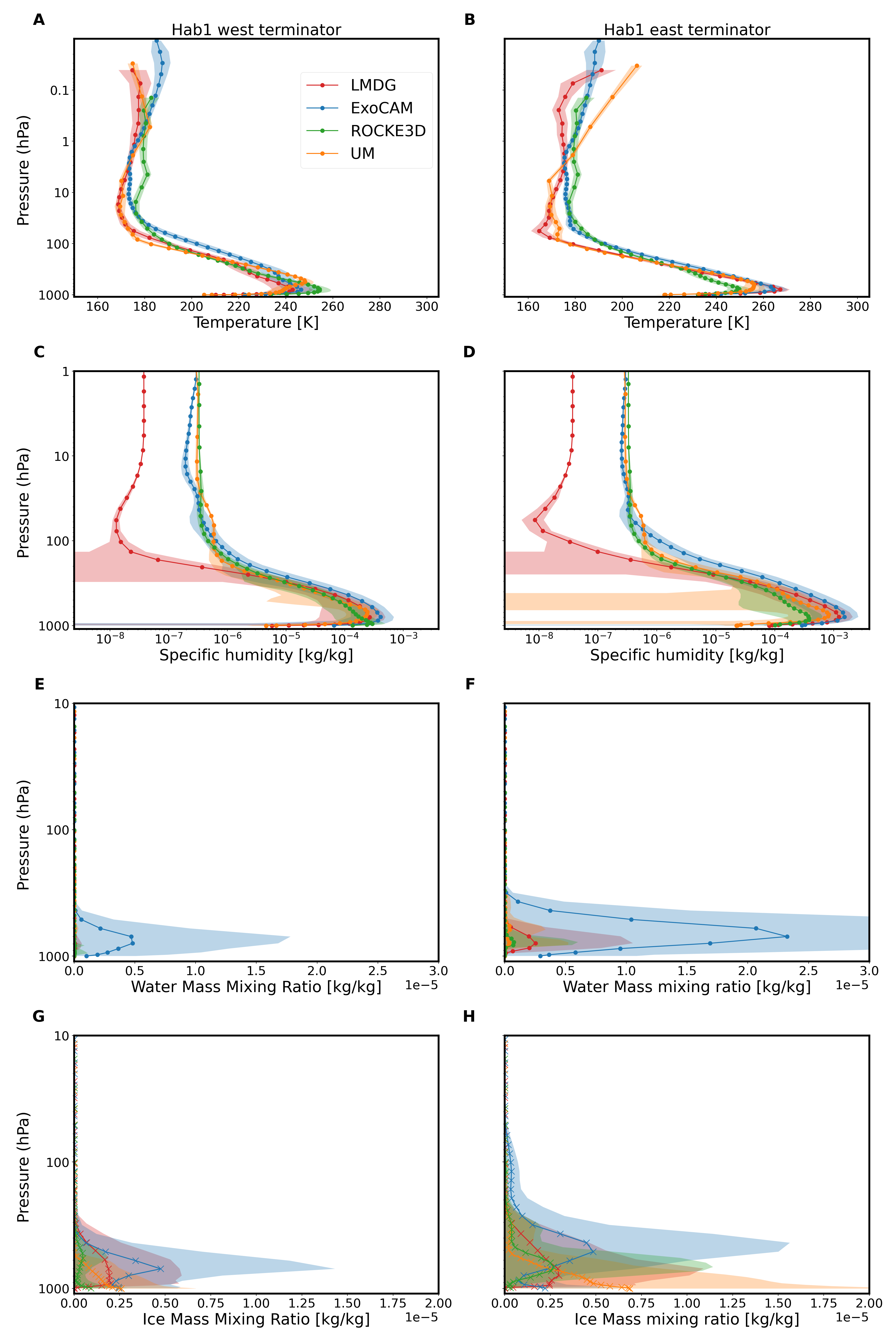}}
\caption{Hab~1 terminator atmospheric profiles predicted by ExoCAM (blue), LMD-G (red), ROCKE-3D (green) and the UM (orange). From top to bottom: temperature, specific humidity, the mass mixing ratio of liquid water and the mass mixing ratio of ice water for the west terminator (left column) and east terminator (right column). Time averaged values are represented by thick lines while the $1-\sigma$ deviations are represented by shades.}
\label{Fig:Hab1_prof}
\end{figure}

Fig.~\ref{Fig:Hab1} shows the Hab~1 transmission spectra calculated using the output from ExoCAM, LMD-G, ROCKE-3D and the UM. The extension of the top of the model has a negligible impact on the spectra, except within the strongest 4.3~$\mu m$ and 15~$\mu m$ CO$_2$ absorption bands. The differences between the spectra from the different GCMs are mostly noticeable in the continuum and for the weakest absorption bands. Indeed, as already shown by \citet{Fauchez:2019} and \citet{Suissa2020}, the continuum level in a cloudy atmosphere is raised to the altitude of the cloud deck. Strong bands like CO$_2$ at 4.3 $\mu m$ are less affected by clouds because even if the denser, most absorbing part of the atmosphere is under them, the efficiency of absorption is so strong that the small CO$_2$ partial pressure remaining above the cloud deck is large enough to saturate the line. The ExoCAM continuum level is the highest, followed by LMD-G, ROCKE-3D and the UM. This is explained by the fact that the liquid water cloud mixing ratio and the altitude of the cloud deck at the west and east terminators, respectively, are much higher in ExoCAM than in LMD-G, ROCKE-3D and the UM, in that order (Fig.~\ref{Fig:Hab1_prof}e,f). Regarding ice clouds, while the mixing ratios predicted by each model are comparable, the average altitude of the clouds is different, with ExoCAM producing the highest ice clouds following by LMD-G, ROCKE-3D and UM (Fig.~\ref{Fig:Hab1_prof}g,h).

Fig.~\ref{Fig:Hab1} also shows that the LMD-G water band around 6~$\mu m$ is significantly weaker than that predicted by other models. This is due to the fact that LMD-G simulations have a much drier upper atmosphere (Fig.~\ref{Fig:Hab1_prof}c,d) . The amount of water above the tropopause in LMD-G is primarily controlled by the tropopause temperature at the substellar point (where water is injected by deep moist convection) which is the coldest in LMD-G, especially at the western terminator \citep[see the companion paper,][Fig.~4]{Sergeev21_THAI}. As a result, the detectability of water in Hab~1~simulations for LMD-G is even more challenging than for the other GCMs.

\begin{figure}[h]
\centering
\resizebox{15cm}{!}{\includegraphics{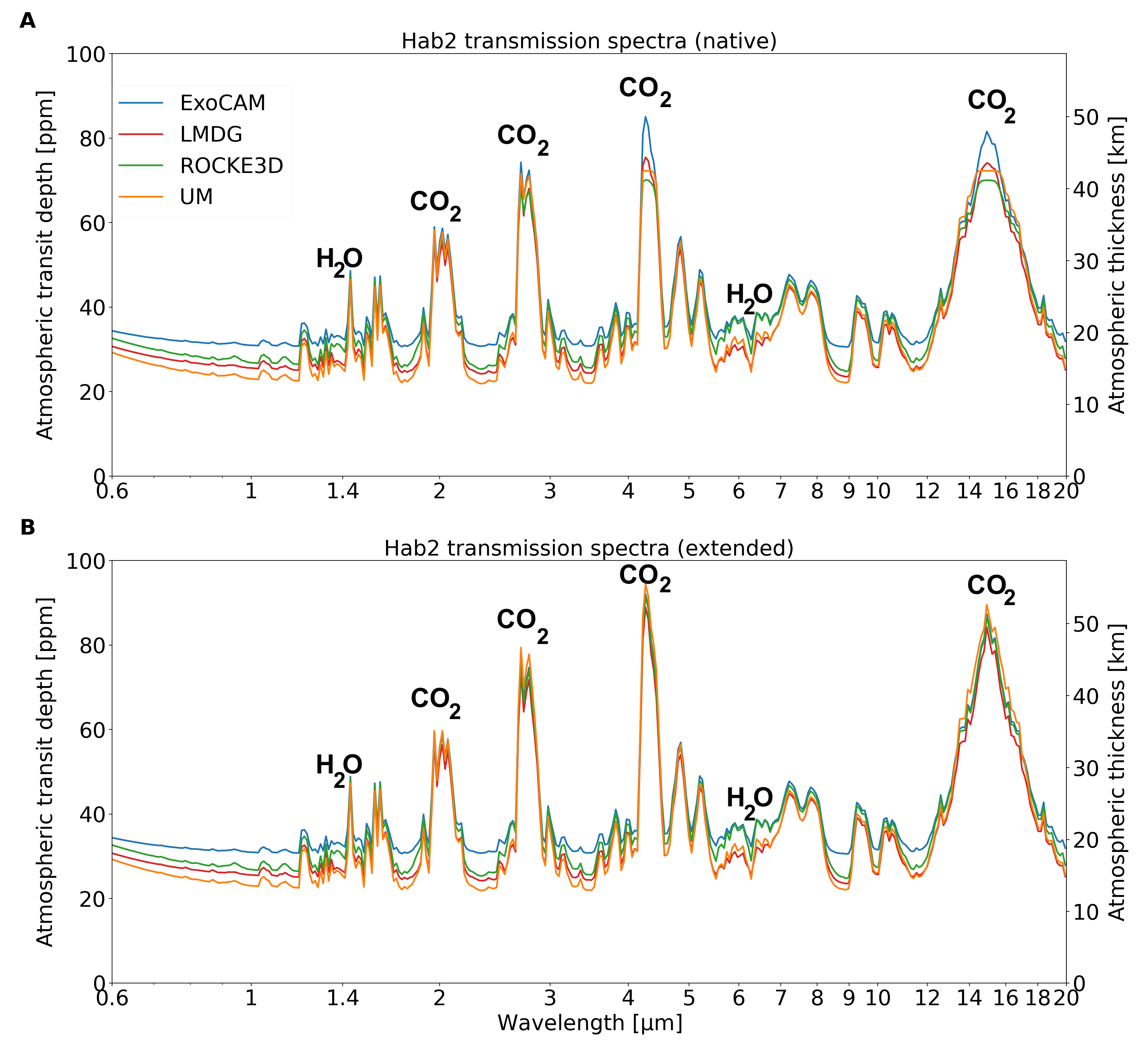}}
\caption{Hab~2 transmission spectra simulated with PSG using the terminator atmospheric profiles from the ExoCAM, LMD-G, ROCKE-3D and the UM simulations. In Panel A) the atmospheric profiles are limited up to the lowest pressure a\added{t} the top of the GCMs, namely 10$^{-5}$, 4$\times$10$^{-5}$, 14$\times$10$^{-5}$ and 13$\times$10$^{-5}$~bar for ExoCAM, LMD-G, ROCKE-3D and UM, respectively  while in Panel B) the atmospheric profiles have been extended up to 10$^{-10}$~bar assuming an isothermal atmosphere and fixed mixing ratios for the dry gases. CO$_2$ features are truncated if the atmospheres are not extended. Differences between the spectra due to clouds are mostly seen in the continuum and for the weakest absorption bands.}
\label{Fig:Hab2}
\end{figure}

Fig.~\ref{Fig:Hab2} is the same as Fig.~\ref{Fig:Hab1}, but for Hab~2 simulations. First, we can see that because the CO$_2$ mixing ratio is much higher in the Hab~2 simulations (from about 400~ppm for Hab~1 to nearly 100~$\%$ for Hab~2), strong absorption lines are more easily truncated by a low model top. Similarly to the Hab~1 case, the ExoCAM continuum is higher than that of the other three GCMs. This time, however, the continuum level in ROCKE-3D is slightly above that in LMD-G. Note that LMD-G's absorption peaks are the smallest among the four GCMs.

\begin{figure}[h]
\centering
\resizebox{15cm}{!}{\includegraphics{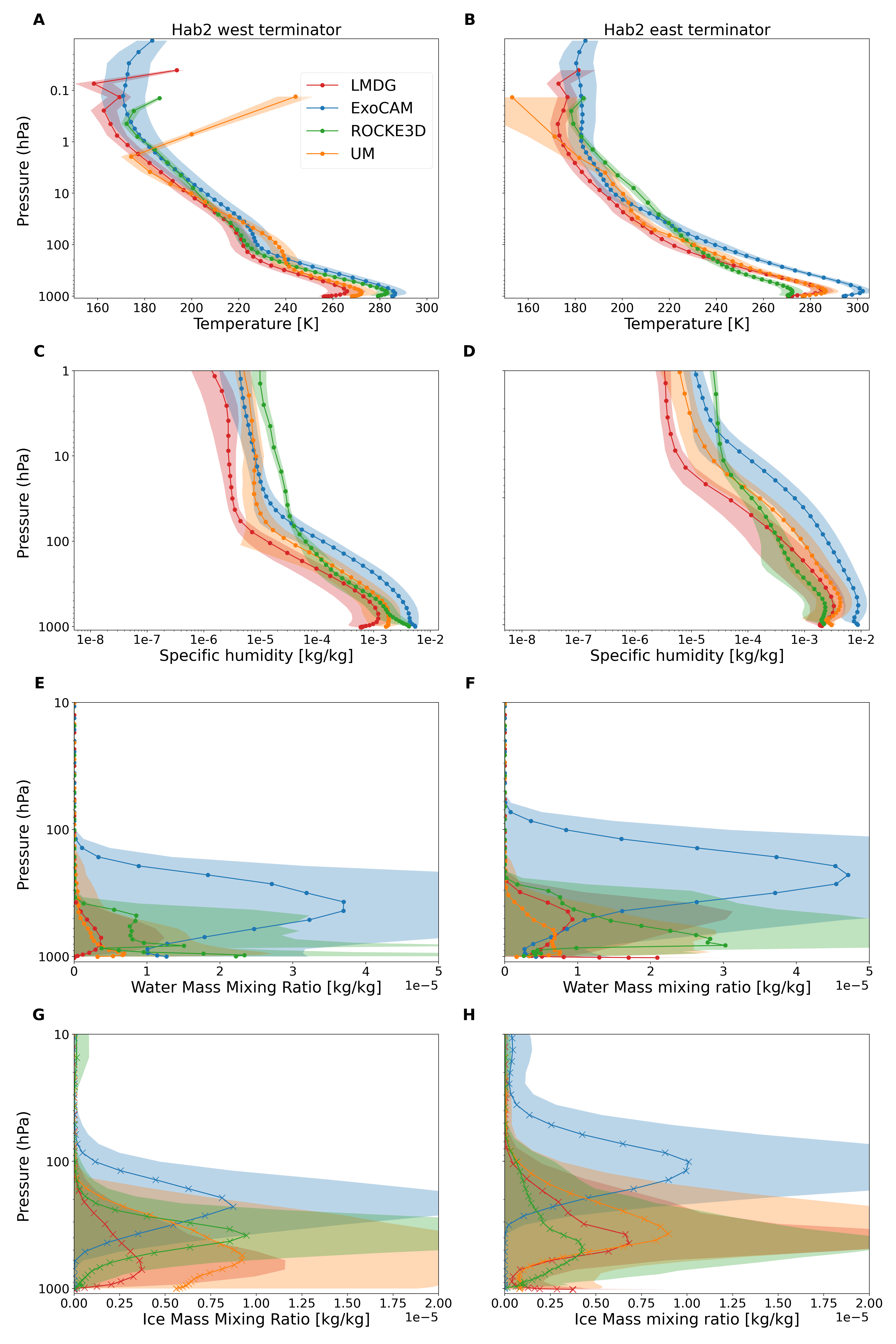}}
\caption{Hab~2 terminator atmospheric profiles simulated by ExoCAM (blue), LMD-G (red), ROCKE-3D (green) and the UM (orange). From top to bottom: temperature, specific humidity, mass mixing ratio of liquid water and mass mixing ratio of ice water for the west terminator (left column) and east terminator (right column). Time averaged values are represented by thick lines while the 1--$\sigma$ deviations are represented by shades.}
\label{Fig:Hab2_prof}
\end{figure}

In the warm and humid atmosphere of the Hab~2 case (Fig.~\ref{Fig:Hab2_prof}a,b) there is no clear temperature inversion at the tropopause except for a decrease in the lapse rate from 100~hPa and lower \citep[for more details see][]{Sergeev21_THAI}. The specific humidity closely follows the temperature profiles: colder temperature profiles correspond to drier profiles (Fig.~\ref{Fig:Hab2_prof}c,d).
Furthermore, the warm atmosphere of Hab~2 results in the liquid water cloud mass mixing ratio being comparable and even larger than the ice cloud mixing ratio (Fig.~\ref{Fig:Hab2_prof}). When the altitudes of both cloud types are combined, ExoCAM has on average higher and thicker clouds, followed by ROCKE-3D, then LMD-G and finally the UM. It is interesting to note that in the warmer, moister and cloudier Hab~2 case, the relative difference in cloudiness between LMD-G, ROCKE-3D and the UM is smaller than that for Hab~1. Only ExoCAM persistent\added{ly} produces higher clouds. More detailed discussion about the differences of cloud coverage produced by the THAI models are given in \citet{Sergeev21_THAI}. 

\begin{figure}[h]
\centering
\resizebox{12cm}{!}{\includegraphics{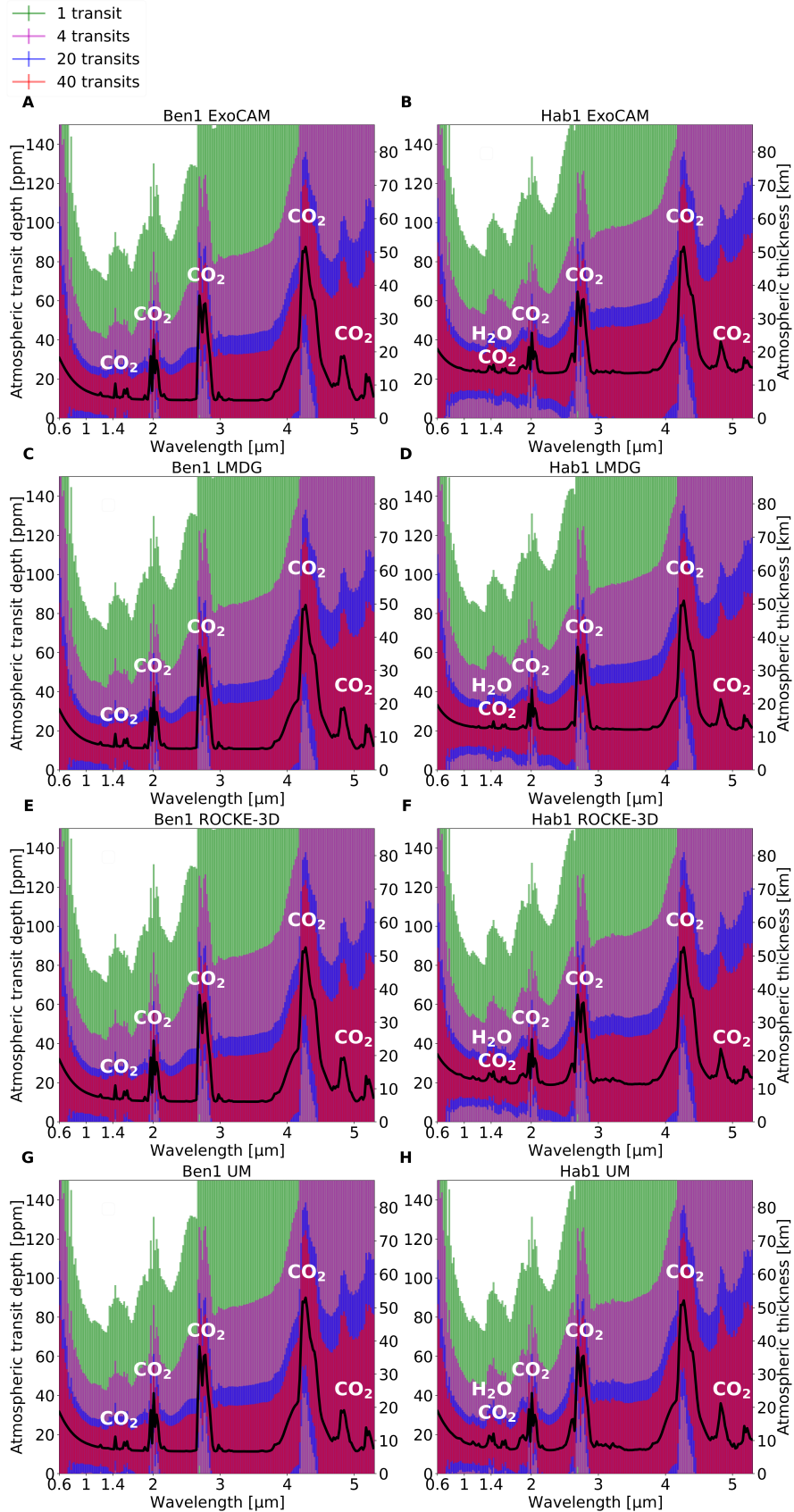}}
\caption{Ben~1 (left column) and Hab~1 (right column) transmission spectra for ExoCAM (A and B panels), LMD-G (C and D panels), ROCKE-3D (E and F panels) and UM (G and H panels). The error bars are shown for the native NIRSpec spectral resolution and for 1 transit (green), 4 transits (magenta, NIRSpec Cycle 1 proposal 1331 \citep{LewisT1eJWST}), 20 transits (blue, $\sim$ 5 JWST years at a constant 4 transits per year) and 40 transits (red, $\sim$ 10 JWST years at a constant 4 transits per year). Relevant absorption features are written in white. }
\label{Fig:Ben1Hab1noise}
\end{figure}

Similar to the Ben~1 \& Ben~2 experiments, we extrapolated the model top for the PSG calculation, which gave the number of transits required for atmospheric detection with a 5--$\sigma$ confidence level by JWST using \added{the full NIRSpec Prism spectral range (5--$\sigma$ Transit) and with} the CO$_2$ line at 4.3$~\mu m$ \added{only (5--$\sigma$ Transit-4.3$\mu m$)}. For Hab~1, we found \replaced{29, 28, 23 and 18}{35, 29, 38 and 23} transits are required\deleted{is the minimum},
while for Hab~2 \replaced{it is}{we found} \replaced{27, 23, 22 and 17}{12, 9, 8 and 7} transits are required (for ExoCAM, LMD-G, ROCKE-3D and UM, respectively). 
\added{The average number of transits for Hab~1 is 29, with a maximum difference of 41$~\%$ and for Hab~2 it is 9, with a maximum difference of 56$~\%$. Similarly to the Ben cases, the use of only the 4.3$~\mu m$ CO$_2$ line overestimates the predicted numbers, especially for the 1~bar CO$_2$ case.} 
\added{Note that the reason why the overall required number of JWST observed transits is higher for the Hab scenarios is the presence of clouds that raise the continuum level up to the cloud deck altitude, shrinking each absorption line from the bottom and therefore reducing their detectability. This can clearly be seen in Fig. \ref{Fig:Ben1Hab1noise} where the left panels show the Ben~1 simulations and the right panels the Hab~1 simulations displaying a change in the continuum level. That change is significantly larger for ExoCAM, which has higher clouds, than UM. We can also see in Fig. \ref{Fig:Ben1Hab1noise} that 1 and 4 transits (green and magenta error bars, respectively) are far from being enough to detect the CO$_2$ features at 5--$\sigma$ but that 20 transits (blue error bars) may be enough for Ben~1, while 40 transits (red error bars) may be necessary for Hab~1. Finally, we can also see that for the Hab~1 cases, water lines around 1.4$~\mu m$ are too small to be detectable, even for the UM showing stronger H$_2$O lines due to its lower cloud deck.}
\deleted{For Hab~1, the difference between the two extremes ExoCAM and and the UM  is about 38$\%$. The difference is slightly lower, 37$\%$ for Hab~2.}

The\deleted{se} differences between the predicted transits \added{from each GCM terminator atmosphere} is to the first order control by the altitude of the cloud deck and to the second order by the temperature profile.  Those differences between ExoCAM and LMD-G on one hand and ROCKE-3D and the UM on the other are significant and could have consequences for the number of hours requested for a JWST proposal and on the interpretation of future data using retrieval algorithms. However, it seems clear that regardless the GCM used to produce the atmospheric data, at least \replaced{17}{7} transit observations would be required to detect at a 5--$\sigma$ confidence level a high molecular weight atmosphere on TRAPPIST-1e with a cloudy sky. \added{Note that during JWST Cycle 1, 4 transits with NIRSpec Prism will be observed as part of the Guaranteed Observation Time (GTO) program 1331 led by PI Nikole Lewis. On average, this would lead to a 2--$\sigma$ detection for Hab~1 and a 3.4--$\sigma$ detection for Hab~2 (see column S/N-4 of Table \ref{tab:transit}).}

\subsection{Inter-transit variability}\label{subsec:variability}
Many previous modeling studies estimating the detectability of an exoplanet through transmission spectroscopy have assumed that each planetary transit will be constant through time \citep{Fauchez:2019,Lustig_Yaeger2019,Komacek2020,Pidhorodetska_2020, Suissa2020,Suissa2020b}. However, this is not a realistic assumption as weather patterns and clouds, if present, are likely to change from one transit to another. Previous work on hot Jupiters by \citet{Komacek&Showman2019} has shown that temporal variability could be already detectable using either secondary eclipse observations with JWST or phase curve observations, and/or Doppler wind speed measurements with high-resolution spectrographs. More recently, \cite{May2021water} simulated TRAPPIST-1e with ExoCAM for various concentration of CO$_2$ and looked at the atmospheric variability between 10 transits induced by ice water clouds. The amplitude of the transit variability for their 10$^{-4}$  bar of CO$_2$ (comparable to Hab~1) and 1~bar CO$_2$ (comparable to Hab~2) are very similar, of the order of 10 and 20~ppm \citep[][their Fig.~4]{May2021water}, respectively. However, they computed one transmission spectrum per day while in our study we compute it at the exact time of the transit, i.e. every 6.1 days, potentially leading to larger atmospheric differences.
The main conclusion is that the time variability of the spectra does not affect retrieved abundances at detectable levels. However, the findings of \citet{May2021water} are likely to be dependent on the GCM (ExoCAM). Here, we analyze the inter-transit variability produced by three more GCMs: LMD-G, ROCKE-3D and the UM.

\begin{figure}[h]
\centering
\resizebox{15cm}{!}{\includegraphics{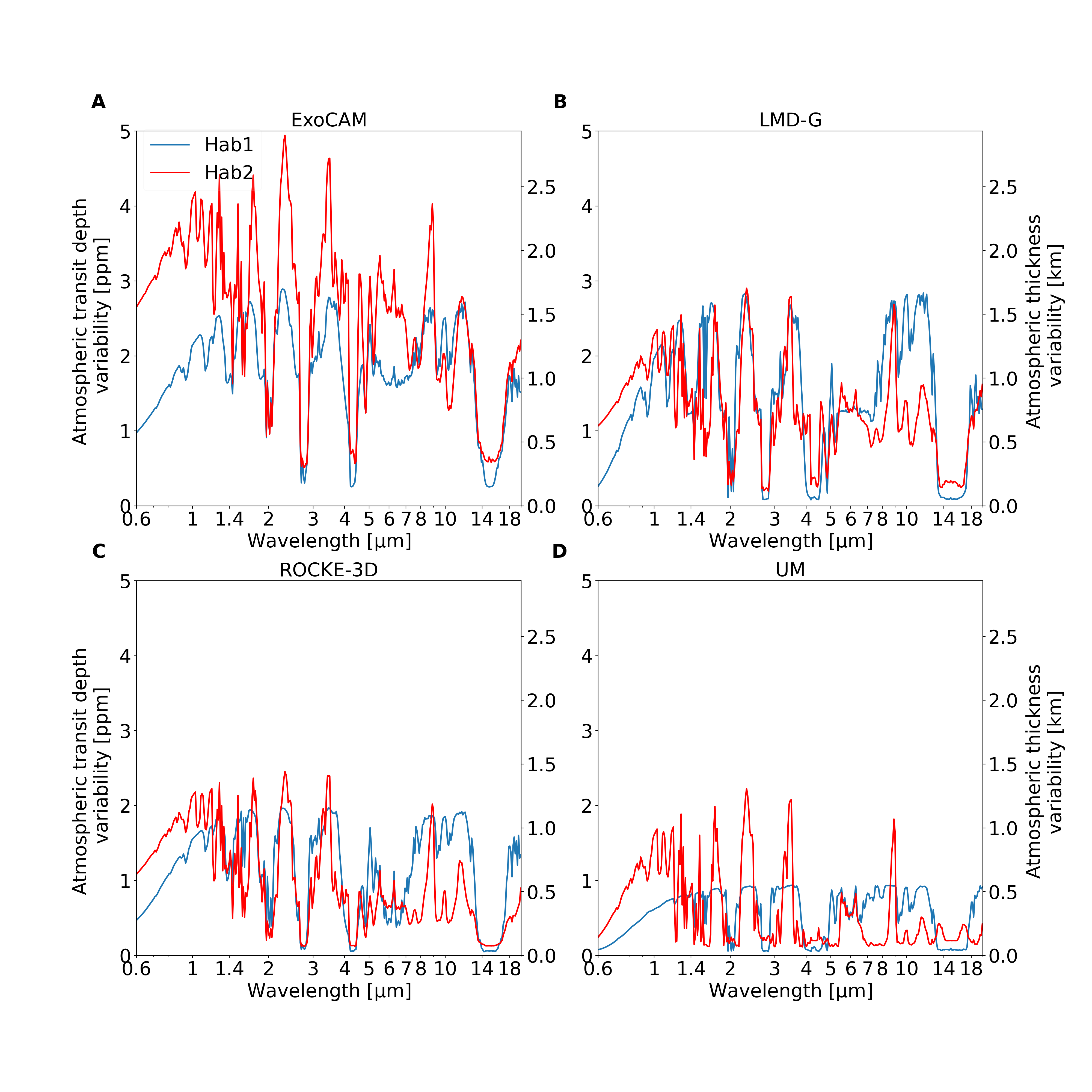}}
\caption{Standard deviation of the atmospheric transit depth [ppm] and of the atmospheric thickness [km] as a function of wavelength for Hab~1 (blue) and Hab~2 (red) and for ExoCAM (panel A), LMD-G (panel B), ROCKE-3D (panel C) and the UM (panel D).}
\label{Fig:varia}
\end{figure}

Fig.~\ref{Fig:varia} shows the standard deviation of the atmospheric transit depth and of the transit atmospheric thickness over 100 transits. This variability is wavelength dependent: it is the largest in the continuum as the transmitted light is closer to the surface, where clouds are present; and the smallest for the strongest absorption lines like the CO$_2$ at 2.7, 4.3 and 15~$\mu m$. There are significant differences between the GCMs. In general\added{,} the cloudier the simulation is, the more variable the transmission spectrum is.
For LMD-G and ROCKE-3D, the time variability in both Hab~1 and Hab~2 is remarkably similar, while for ExoCAM and the UM it differs. This difference is due to the change in the average altitude of clouds between Hab~1 (\added{Figure} \ref{Fig:Hab1_prof}) and Hab~2 (\added{Figure} \ref{Fig:Hab2_prof}). In LMD-G and ROCKE-3D simulations, the average altitude of liquid water and ice water clouds does not change substantially between Hab~1 and Hab~2, while for ExoCAM and the UM it increases sharply in Hab~2. For instance, for ExoCAM the east terminator water ice clouds maximum density peaks at 500~hPa for Hab~1 and at 250~hPa for Hab~2. We hypothesise that stronger winds at this lower pressure relative to those deeper in the atmosphere lead to higher cloud variability. 
Overall, the standard deviation of the continuum level in the Hab~1 case for ExoCAM, LMD-G, ROCKE-3D and the UM is about 3, 3, 2 and 1~ppm, respectively, leading to a median value of $\sim$2~ppm. For Hab~2 it is about 5, 3, 2.5 and 2~ppm, respectively, leading to a median value of $\sim$3~ppm. These values are low relative to the JWST expected 1--$\sigma$ noise of 10 to 25~ppm \added{as assumed in} \cite{Fauchez:2019} \added{and near the upper limit value (14~ppm) estimated by NIRSpec lab time series in \cite{Rustamkulov2022}}. It is also interesting to note that those values are comparable to the relative transit depth of H$_2$O or O$_2$ \citep{Fauchez:2019,Lustig_Yaeger2019,Pierrehumbert2010,Wunderlich2020}. This means that even if one assume\added{s} no noise floor, atmospheric variability would produce a continuum fluctuation that would swamp those highly important but weak absorption lines.

\begin{figure}[h]
\centering
\resizebox{15cm}{!}{\includegraphics{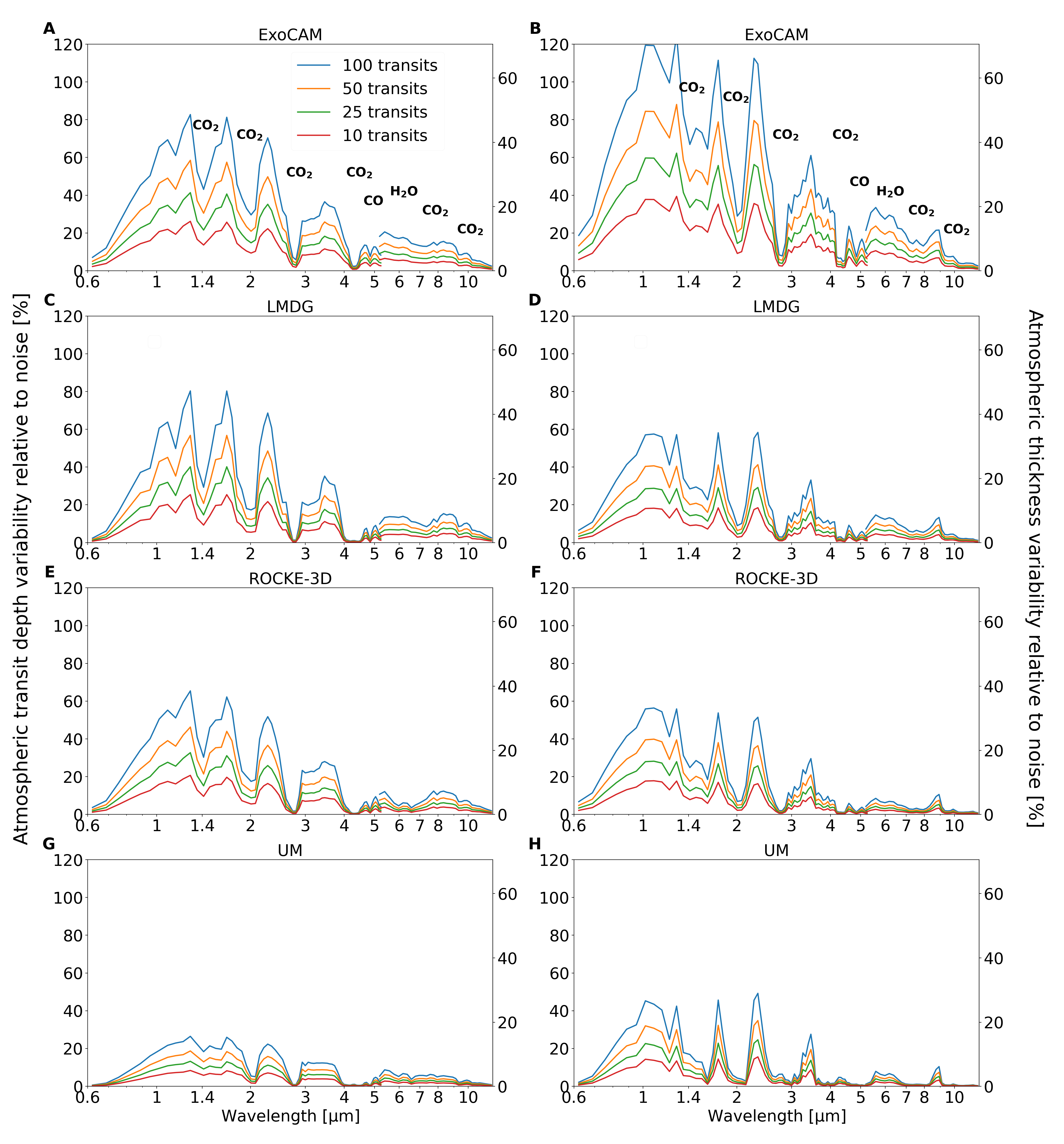}}
\caption{Atmospheric transit depth variability relative to noise (left Y axis) and atmospheric thickness variability relative to noise (right Y axis) for 100 (blue), 50 (orange), 25 (green) and 10 (red) transits for ExoCAM (panels A and B), LMD-G (panels C and D), ROCKE-3D (panels E and F) and the UM (panels G and H). Hab~1 is on the left column, Hab~2 on the right column. The noise is computed at the native instrument resolution and later binned down by a factor 3 for this figure.}
\label{Fig:specvar}
\end{figure}

Fig.~\replaced{\ref{Fig:varia}}{\ref{Fig:specvar}} shows the spectra as the ratio in percentage of the standard deviation of the variability with respect to the measurement noise for 100 (blue), 50 (orange), 25 (green) and 10 (red) transits for the atmospheric transit depth (left Y axis) and atmospheric thickness (right Y axis). The larger the number of transits, the lower the noise and therefore the higher the variability-to-noise ratio (\%).
Interestingly, these spectra can be reminiscent of emission spectra \citep{Morley2017,Lustig_Yaeger2019,Fauchez:2019}. The minimum values correspond to the absorption line peak, while the maxima correspond to the continuum.  Only the Hab~2 atmosphere simulated by ExoCAM would lead to a transit depth and atmospheric thickness variability higher than the measurements noise if 100 transits or more are acquired with JWST. \added{Considering observational constraints and science priority this number is likely too high. \cite{Fauchez:2019} have used the \url{https://exoctk.stsci.edu/contam_visibility} tool to estimate the number of times TRAPPIST-1e will be visible transiting over JWST 5 years of nominal life time and found 85 (17 transit per year).  The recent successful launch of JWST and optimized Ariane V trajectory saving large amounts of fuel has allowed us to extend JWST's potential lifetime up to 20 years. If every single transit is effectively observed this is an upper limit of 340 available transits. However for Cycle 1, only 4 transits are going to be observed, if we assume that number constant over 20 years it will bring up to 80 transits. In our work, we have therefore considered 50 non-consecutive transits as a realistic but conservative estimate} In a more realistic scenario of 50 non-consecutive transits accumulated over the lifetime of the JWST, the impact of atmospheric variability relative to the noise for Hab~1 and Hab~2 would be of about 50$\%$ and 80$\%$ for ExoCAM, 50$\%$ and 50$\%$ for LMD-G\deleted{, for Hab~1 and Hab~2}, 40$\%$ and 40$\%$ for ROCKE-3D and 20$\%$ and 40$\%$ for the UM, respectively, and will therefore \added{not} be of a concern.

\begin{figure}[h]
\centering
\resizebox{15cm}{!}{\includegraphics{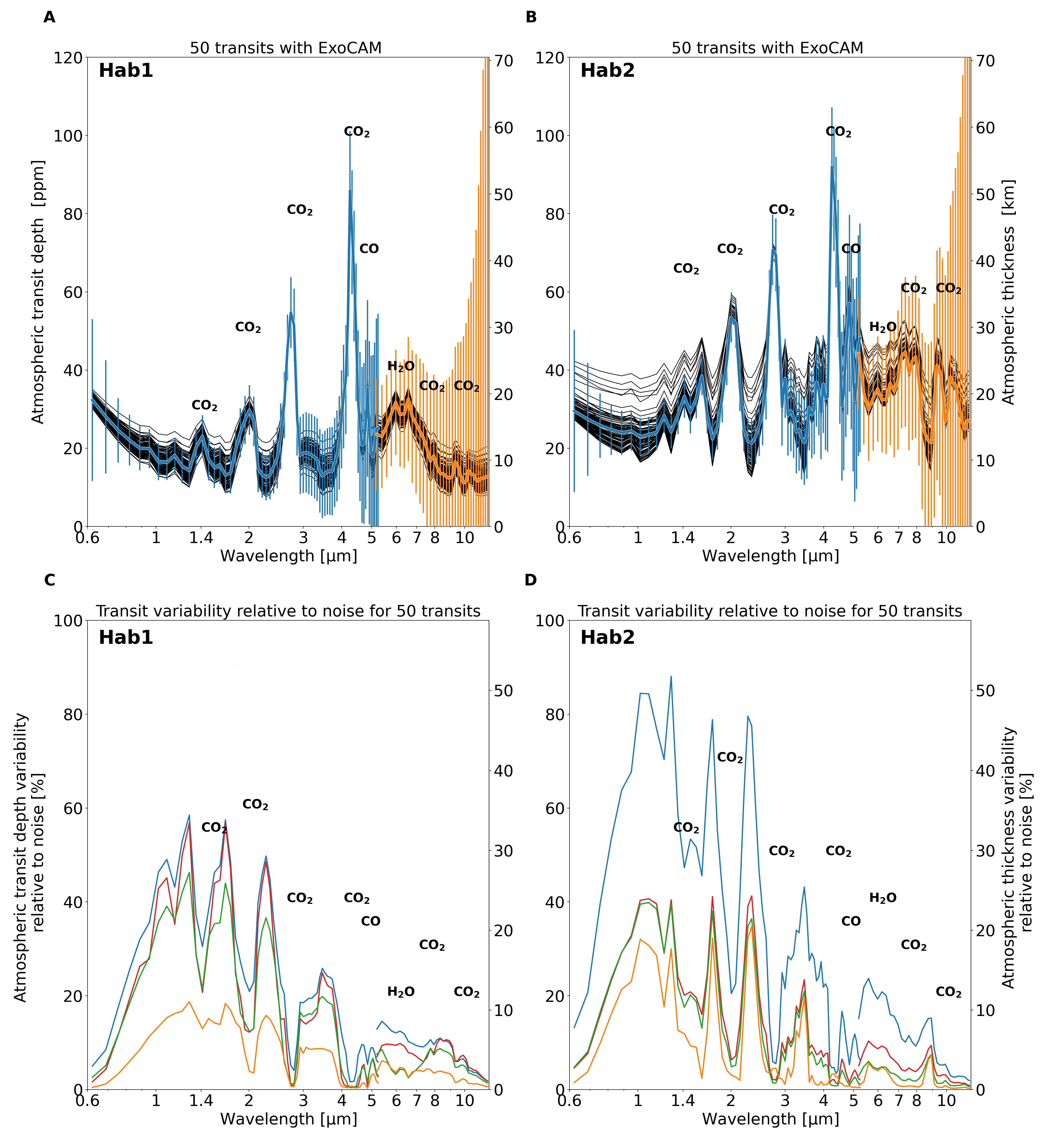}}
\caption{Panels A and B: transmission spectra for Hab~1 and Hab~2, respectively, obtained from ExoCAM simulations. Each fine black trace represents 1 of the 100 individual transits and the thick blue line represent the median transit. The noise for 50 transits and for both NIRSpec Prism (blue) and MIRI (orange) has been binned down by a factor 3 to maximize the number of photons per spectral bin while preserving the line shapes and  is represented by the vertical error bars. Panels C and D: Hab~1 and Hab~2, respectively, atmospheric transit depth variability relative to noise (left Y axis) and atmospheric thickness variability relative to noise (right Y axis) for 50 transits and each of the four GCMs.}
\label{Fig:specvar2}
\end{figure}

Comparisons between all GCMs in a single panel is shown in Fig. \replaced{\ref{Fig:specvar}}{\ref{Fig:specvar2}}. In panels A (Hab~1) and B (Hab~2) are shown each ExoCAM transmission spectrum (black lines) and the median spectrum (blue line) with the associated 1--sigma error (blue error bars). ExoCAM was selected for this example as it is the GCM showing the largest variability. In panels C (Hab~1) and D (Hab~2) the spectra of the variability relative to noise for 50 transits and for the four GCMs are shown.  

As a summary, in the case of TRAPPIST-1e it seems predictions of the atmospheric variability introduced by clouds for an N$_2$ or CO$_2$-dominated atmosphere are within the measurements noise for a reasonable number of transits ($<$~50) regardless of the GCM used to simulate this temporal and spatial variability. However, if in the fortunate event \citep[albeit unlikely; see the discussion in][]{gillon2020trappist1} that a similar exoplanet were to be found closer \added{to Earth}, the noise will be reduced. In that case the atmospheric variability could be a possible proxy for the presence of clouds via the temporal changes of the continuum level relative to the relatively stationary strong absorption peaks. Without such variability, a continuum level corresponding to a cloud deck, an atmospheric refraction limit or a planet's surface may not be discernible. Additionally, ExoCAM simulations produce systematically higher variability in the synthetic spectra compared to that in the three other models, while the UM tends to produce  the lowest variability. LMD-G and ROCKE-3D are in the middle. We recommend that these differences in variability are taken into account when using a single GCM for future studies. Finally, \cite{Sergeev21_THAI} have noticed that the time scale of this periodic variability differs between the models (see their Fig. 12). The tendency being the same for Hab~1 and Hab~2 with ExoCAM showing the longest period of 12.5 orbits for both cases, followed by LMD-G with 11.1 and 7.7 orbits, respectively, ROCKE-3D with 2.3 and 3 orbits, respectively and UM with 1.1 and 2.5 orbits, respectively.  With real atmospheric data it is unknown what this period would be within this 1 to 12.5 orbit range. As a result, it is not clear if observing consecutive transits or scattered one\added{s} would have any impact on the atmospheric characterization.
Finally, it is worth noting that while TRAPPIST-1e lies between the fast and Rhines rotation regimes \citep{Turbet21_THAI, Sergeev21_THAI}, planets with longer orbital periods that remain in synchronous rotation will transition from the Rhines rotation regime to a slow rotation regime, which increases the symmetry of temperature and winds about the substellar point \citep{Haqq_Misra2018}. However, how the atmospheric variability across transits would be affected by the change of atmospheric regime remains to be explored.

\section{Conclusions} \label{sec:conclusions}
In this third and last part of the THAI paper series we analyzed how the prediction of the detectability of N$_2$ and CO$_2$-dominated atmospheres on TRAPPIST-1e is sensitive to the choice of a 3D GCM used to simulate its atmosphere.

% MORE GENERAL CONCLUSIONS ABOUT THAI TBD WITH PART I and PART II CONCLUSIONS. Martin and Denis
%In the Part I \citep{Turbet21_THAI} of this work, we have compared the atmospheres and surface properties produced by the four GCMs ExoCAM, LMDG, ROCKE-3D and the UM for the Ben~1 \& Ben~2 scenarios consisting of modern Earth and CO$_2$ -like atmospheres with no water in the atmosphere nor on the surface. Except small differences in atmospheric circulation and upper atmosphere temperature, the models agree quite well, with average surface temperatures varying of only few degrees.
First, we have simulated the transmission spectra for the  Ben~1 \& Ben~2 scenarios (dry land planets for which the comparison of the predicted atmosphere is presented in Part I \cite{Turbet21_THAI}) using the Planetary Spectrum Generator (PSG, \cite{Villanueva2018}).
We have shown that the predicted spectra are similar between the GCMs and the number of transits to detect an atmosphere with a 5--$\sigma$ confidence level is \replaced{almost identical}{close} the for Ben~1 (within \replaced{1}{4} transits or 24$~\%$) and Ben~2 (within \replaced{3}{2} transits or 33$~\%$) cases.

Concerning the aquaplanet scenarios, presented in Part II \citep{Sergeev21_THAI} of this work, we have shown that in the terminator region, in the Hab~1 case, the ExoCAM water cloud mixing ratio (only true for liquid water) and altitude of the cloud deck are much higher than for LMD-G, ROCKE-3D and the UM, in that order.  In the Hab~2 case, ExoCAM has on average higher thick clouds, followed by ROCKE-3D, and closely by LMD-G and finally the UM, with the relative difference between the three latter models being smaller than for the Hab~1 case. The large cloud mixing ratio for LMD-G was at first counter-intuitive as it employs a convective adjustment that notoriously produces fewer convective clouds than the mass-flux scheme used by the three other models, as also shown in \citep{Yang2013,Sergeev21_THAI}. Here, at the terminator, the fact that LMD-G is cloudier than ROCKE-3D and UM is likely due to a larger production of stratiform clouds. Unfortunately, the THAI protocol did not include separate output for convective and stratiform clouds. This differentiation will be investigated in a future study.

The differences in the simulated cloud coverage between the models, along with changes in temperature and water vapor profile lead to  \replaced{$\sim~$30--40$\%$}{41 to 56$\%$} differences between the number of transits predicted to detect (at 5--$\sigma$) molecular species \added{in Hab~1 and Hab~2 cases, respectively,} using the atmospheric profile of a GCM or another. These differences are non-negligible as they can change about $1-\sigma$ the confidence level of a predicted detectability of an atmosphere or increase the observing time by \replaced{$\sim$30--40$\%$}{41 to 56$\%$}, potentially making a given observation proposal unfeasible. Without observational data, we do not know which model is closer to the truth, but comparing them against each other indicates whether the detectability estimate is optimistic or pessimistic. This work therefore provides for the first time a ``GCM uncertainty error bar'' of \replaced{$\sim$30--40$\%$}{$\sim 50\%$} that needs to be considered in future analysis of JWST spectra of TRAPPIST-1e. Namely, simulations of temperate rocky exoplanets with ExoCAM would likely produce higher and thicker clouds relative to other GCMs. As a result, a tool like PSG would give a higher number of transits required to detect such an atmosphere.
On the other hand, using ROCKE-3D or the UM would give a lower number of transits, because they would likely produce lower and thinner clouds. As for LMD-G, the number would be comparable to that of ExoCAM due to LMD-G's colder upper atmosphere.
\added{It also seems that these relative differences increase with the atmospheric concentration of the gas (here CO$_2$) that is being retrieved. This is because more minor lines appear and as they are shallow they are more sensitive to the cloud properties predicted by a given GCM. However, a larger gas concentration generally lead to an lower average absolute number of transits.}
Note that with the 4 transits expected for NIRSpec Prism in the JWST Guaranteed Observation Time (GTO, program 1331, PI Ni\replaced{c}{k}ole Lewis), we can expect \replaced{at best a 2.5--$\sigma$  confidence level if the atmosphere is Earth-like with a similar amount of clouds than predicted by ROCKE-3D or UM.}{an average of 2.6 and 4.3~--$\sigma$ for the dry atmospheres Ben~1 and Ben~2 and an average of 2.0 and 3.4 ~$\sigma$ for the moist and cloudy Hab~1 and Hab~2 atmospheres, respectively.}

% ADD BEYOND THAI -> CUISINES
THAI has been well received by the community, as demonstrated by the attendance of 125 people at the THAI workshop and the 35 authors of the THAI workshop report \citep{Fauchez2021_THAI_workshop}. Due to extreme paucity of observational data it is important for the exoplanet community to develop and maintain intercomparison frameworks to benchmark atmospheric models, improve physical parameterizations and evaluate their sensitivity. In this context, THAI is the first step toward a larger framework of intercomparison for exoplanets, the Climates Using Interactive Suites of Intercomparisons Nested for Exoplanet Studies (CUISINES). Within the CUISINES framework, we hope to develop intercomparison projects similar to THAI for exoplanets other than TRAPPIST-1e using an hierarchy of numerical models. Ultimately, the goal of CUISINES is to provide the exoplanet community, both on the modelling and observational ends of the spectrum, with model benchmarks and recommendations for comparison with existing observations and for planning future ones.

\acknowledgments
\added{The authors thank the two anonymous reviewers whose comments helped to improve the quality and clarity of this manuscript.}
T.J.F., G.L.V. and M.J.W. acknowledge support from the GSFC Sellers Exoplanet Environments Collaboration (SEEC), which is funded in part by the NASA Planetary Science Divisions Internal Scientist Funding Model.
D.E.S., I.A.B., J.M. and N.J.M. acknowledge use of the Monsoon system, a collaborative facility supplied under the Joint Weather and Climate Research Programme, a strategic partnership between the Met Office and the Natural Environment Research Council.
We acknowledge support of the Met Office Academic Partnership secondment program.
This work was partly supported by a Science and Technology Facilities Council Consolidated Grant (ST/R000395/1), UKRI Future Leaders Fellowship (MR/T040866/1), and the Leverhulme Trust (RPG-2020-82).\\
This project has received funding from the European Union's Horizon 2020 research and innovation program under the Marie Sklodowska-Curie Grant Agreement No. 832738/ESCAPE.
M.T. thanks the Gruber Foundation for its generous support to this research.
\added{M.T. acknowledges support from the PORTAL BRAIN-be 2.0 BELSPO project.}
M.T. was granted access to the High-Performance Computing (HPC) resources of Centre Informatique National de l'Enseignement Sup\'erieur (CINES) under the allocations \textnumero~A0020101167 and A0040110391 made by Grand \'Equipement National de Calcul Intensif (GENCI).
This work has been carried out within the framework of the National Centre of Competence in Research PlanetS supported by the Swiss National Science Foundation.
M.T. acknowledges the financial support of the SNSF. M.T. and F.F. thank the LMD Generic Global Climate team for the teamwork development and improvement of the model. 
J.H.M. acknowledges funding from the NASA Habitable Worlds program under award 80NSSC20K0230.\\
The authors acknowledge the help of Andrew Ackerman to set up the cloud diagnostics in ROCKE-3D.\\
\added{T.F. thank Avi. Mandell for his helpful discussion on the observation noise.}
The THAI GCM intercomparison team is grateful to the Anong's Thai Cuisine restaurant in Laramie for hosting its first meeting on November 15, 2017.\\
Numerical experiments performed for this study required the use of supercomputers, which are energy intensive facilities and thus have non-negligible greenhouse gas emissions associated with them.

%% To help institutions obtain information on the effectiveness of their 
%% telescopes the AAS Journals has created a group of keywords for telescope 
%% facilities.
%
%% Following the acknowledgments section, use the following syntax and the
%% \facility{} or \facilities{} macros to list the keywords of facilities used 
%% in the research for the paper.  Each keyword is check against the master 
%% list during copy editing.  Individual instruments can be provided in 
%% parentheses, after the keyword, but they are not verified.

\software{{\sc matplotlib} \citep{Hunter2007}.
PSG \citep{Villanueva2018} is available online at \url{https://psg.gsfc.nasa.gov/index.php}.
 ExoCAM \citep{Wolf2015} is available on Github: https://github.com/storyofthewolf/ExoCAM. The Met Office Unified Model is available for use under licence; see \url{http://www.metoffice.gov.uk/research/modelling-systems/unified-model}. ROCKE-3D is public domain software and available for download for free from \url{https://simplex.giss.nasa.gov/gcm/ROCKE-3D/}. Annual tutorials for new users take place annually, whose recordings are freely available on line at \url{https://www.youtube.com/user/NASAGISStv/playlists?view=50&sort=dd&shelf_id=15}. LMD-G is available upon request from Martin Turbet (martin.turbet@lmd.jussieu.fr) and François Forget (francois.forget@lmd.jussieu.fr).         }

%% Appendix material should be preceded with a single \appendix command.
%% There should be a \section command for each appendix. Mark appendix
%% subsections with the same markup you use in the main body of the paper.

%% Each Appendix (indicated with \section) will be lettered A, B, C, etc.
%% The equation counter will reset when it encounters the \appendix
%% command and will number appendix equations (A1), (A2), etc. The
%% Figure and Table counter will not reset.

\appendix

\section{Data accessibility}
All our GCM THAI data are permanently available for download here: \url{https://ckan.emac.gsfc.nasa.gov/organization/thai}, with variables described for each dataset. If you use those data please cite this current paper and add the following statement: "THAI data have been obtained from \url{https://ckan.emac.gsfc.nasa.gov/organization/thai}, a data repository of the Sellers Exoplanet Environments Collaboration (SEEC), which is funded in part by the NASA Planetary Science Divisions Internal Scientist Funding Model."\\
Scripts to process the THAI data are available on GitHub: \url{https://github.com/projectcuisines}\\
Scripts to generate PSG/GlobES spectra are available on GitHub: \url{https://github.com/nasapsg/globes}.

\bibliography{biblio}{}
\bibliographystyle{aasjournal}

%% This command is needed to show the entire author+affiliation list when
%% the collaboration and author truncation commands are used.  It has to
%% go at the end of the manuscript.
%\allauthors

%% Include this line if you are using the \added, \replaced, \deleted
%% commands to see a summary list of all changes at the end of the article.
%\listofchanges

\end{document}